\documentclass[nolinenumbers]{aastex631}
\newcommand{\chletg}{{\small {\it Chandra}/LETG}}
\newcommand{\sco}{Nova Sco 2023}
\newcommand{\xrgs}{{\small {\it XMM-Newton}/RGS}} 

\begin{document}

\title{X-ray observations of \sco: Spectroscopic evidence of charge exchange}

\correspondingauthor{Sharon Mitrani}
\email{sharonm@campus.technion.ac.il}

\author{Sharon Mitrani}
\affiliation{Department of Physics, Technion, Haifa 32000, Israel}

\author{Ehud Behar}
\affiliation{Department of Physics, Technion, Haifa 32000, Israel}

\author{Marina Orio}
\affiliation{Department of Astronomy, University of Wisconsin, 475 N. Charter Str., Madison, WI 53706, USA}
\affiliation{INAF-Padova, vicolo Osservatorio 5, I-35122 Padova, Italy}

\author{Jack Worley}
\affiliation{Department of Astronomy, University of Wisconsin, 475 N. Charter Str., Madison, WI 53706, USA}




\begin{abstract}
The super-soft source (SSS) phase of a nova eruption, observed a few days after the outburst,  usually
displays an absorbed X-ray thermal continuum with absorption features, emitted by the white dwarf (WD) atmosphere. However, the X-ray spectra of 
many novae in this phase display additional emission lines which likely originate from shocks in the novae ejecta. When the shocked plasma interacts with cold gas, narrow radiative recombination continua (RRCs) and charge exchange (CX) emission are observed. We present the analysis of high-resolution \chletg\ X-ray grating spectra of \sco, observed 128 and 183 days after the optical peak, on 2023 August and October. At both epochs, the absorbed X-ray thermal continuum is well described by a 
Non-Local Thermal Equilibrium atmosphere model with a temperature $T\simeq$ 750,000 K ($kT\simeq$ 65 eV).
On day 128, the atmosphere is found to be outflowing at $v\simeq-3500$\,km\,s$^{-1}$. On day 183, the atmosphere brightened by a factor of $\sim2$ and slowed down to $v\simeq-1500$\,km\,s$^{-1}$.
The discrete emission features of the spectrum consist of the C$^{+5}$, N$^{+5}$, and N$^{+6}$ RRCs, indicating a cold electron temperature of $kT_e\simeq1$\,eV on day 128, and $kT_e\simeq20$\,eV on day 183. The observed line series of H-like and He-like C$^{+5}$, N$^{+5}$, N$^{+6}$, and O$^{+6}$ show enhanced intensities of high-$n$ (principal quantum number) transitions, consistent with a CX model of hot ions at $kT \sim 100$\,eV. The velocity shift of the CX lines remained at $v\sim\pm3000$\,km\,s$^{-1}$, which can be explained by a bipolar outflow.
After Nova Ret 2020 (YZ Ret), \sco\ is yet another nova in which we have found exquisite evidence of CX in astrophysical ionized plasma. 
\end{abstract}



\section{Introduction} \label{sec:intro}
Novae are thermonuclear runaway eruptions on the outer layers of white dwarfs (WDs) following accretion events from a stellar companion in a close, interacting binary. After the outburst,  the WD atmosphere contracts at constant bolometric luminosity, the effective temperature increases and the peak wavelength of the emission moves from the optical range, all the way to the soft X-rays range within  days to months. The central source appears as a thermal continuum X-ray supersoft source (SSS), in which we observe a supersoft X-ray flux below 0.8\,keV \citep[see review by][]{Orio2012}. 

In this SSS phase, some novae show an X-ray continuum with absorption features (absorption lines, ionization edges), some show emission spectra, and some show a combination of both. \citet{Ness2013} suggest a connection between the obscuration of the system and two spectral types: SSa (absorption) or SSe (emission). It is likely that in low inclination systems, the WD atmosphere can be observed directly, with its typical absorption features (SSa). For high inclination, the WD atmosphere can be obscured by a disk or the surrounding ejecta, thus the emitted spectrum is dominated by emission features (SSe). 

However, many novae have intermediate inclination, so they display the SS continuum with superimposed emission features. The emission features can also be present independently from the inclination, because they arise in powerful shock in colliding winds at different velocity, and these can occur even after months or far from the WD \citep[e.g. RS Oph]{Nelson2008, Ness2023}. These papers and several more that analyse the initial phase, when the WD continuum has not appeared yet \citep[e.g.][]{Peretz2016, Orio2020, Orio2022} describe how the emission lines can be ascribed to collisionless shocked gas. Novae thus provide an ideal laboratory for studying shocks, which heat the gas to X-ray temperatures (several keV). The shocked outflows of hot, ionized plasma travel at high velocities of 1000s km\,s$^{-1}$ and eventually collide with ambient cold gas. The most significant evidence of this interaction is charge exchange (CX) between highly ionized species and neutral atoms. For the first time in a nova, evidence for CX has been reported for Nova Ret 2020 (YZ Ret) \citep{Mitrani2024}.

  The target of our investigation, Nova Scorpii 2023 (also referred as V1716 Sco), which is located at a distance of a few kpc \citep{Wang2024},  was discovered on April 4, 2023 by optical observations \citep{Izzo2023}. It was then immediately observed at all wavelengths, including infrared \citep{Woodward2024}, X-rays \citep{Sokolovsky2023}, and $\gamma$-rays \citep{Cheung2023}. In X-rays, the nova was monitored with {\sl Swift} during most of the outburst. The SSS phase began around 6 weeks after the optical outburst \citep{Wang2024}, and was also monitored by {\sl NICER} for almost 2 months. Further results of the X-ray monitoring are reported in an accompanying paper (Worley et al., submitted).
   We observed \sco\ with {\sl Chandra} on 2023 August 25 and 26 (twice) and between  2023 October 20 and 22 (three times).
 The accompanying paper (Worley et al.) analyzes the
lightcurves and variability properties of the nova, finding
 among other features a periodic oscillation with a 79.6 s period that they suggest is the WD spin period. 
 The continua in the X-ray high resolution  grating spectra of the nova were also examined and revealed the presence of a WD atmosphere, most likely at a temperature of $\simeq$750,000 K, in 2023 October (day 183), but the superimposed emission features could not be fitted with a model of thermal plasma in collisional ionization equilibrium \citep['\textit{apec}' in XSPEC,][]{Smith2001} like in several other novae in the SSS phase \citep[e.g.][]{Drake2016, Peretz2016, Orio2020}. In this paper, we start from this point and show how to interpret these puzzling spectra.

 
\section{Observations and data reduction} \label{sec:obs}
The first run of {\sl Chandra} observations started 128 days after the optical peak (in August), with two exposures, the second run was done after 183 days (in October),
with three exposures. Low Energy Transmission Grating (LETG) spectra were obtained with the High Resolution Camera (HRC) aboard {\sl Chandra}. Details of the exposures, including count rates and flux in the 15-35\AA\ wavelength range, are reported in Table \ref{tab:observations}.
We extracted the five spectra using the CIAO software package \citep{Fruscione2006} with CALDB version 4.10.4. The \texttt{chandra\_repro} tool was 
applied to the event file, then we averaged the +1 and -1 diffraction orders with the \texttt{combine\_grating\_spectra} tool.
\
We analyse here the LETG spectra
in the wavelength range of 15–32 \AA, corresponding to the 0.38–0.83 keV energy range. Outside this range, there are only a few counts, therefore we need not worry about higher diffraction orders. Above 32\AA, the source is absorbed by the interstellar medium (ISM). The two spectra observed on days 128 and 129 show only minor differences, while the three spectra observed on days 183, 184, and 185 have practically no difference. Thus, the spectra of each epoch were fitted simultaneously. The two epochs will be referred to as 'd128' and 'd183'.

\begin{deluxetable}{cccccc}
\tablecaption{Summary of {\sl Chandra} X-Ray Observations: date, average
 count rate and flux in the 15-35 \AA\ range. \label{tab:observations}}
\tablewidth{0pt}
\tablehead{
\colhead{\textbf{Obs. ID}}  & \colhead{\textbf{Obs. Date}} & \colhead{\textbf{Day Post-Outburst}} & \colhead{\textbf{Exposure  (s)}} & \colhead{\textbf{Count Rate (ct \(\mathrm{s}^{-1}\))}} &
\colhead{\textbf{Flux $\times 10^{-11} \, \mathrm{erg} \, \mathrm{cm}^{-2} \, \mathrm{s}^{-1}$}}
}
\startdata
28048 & 2023 Aug 26 & 128 & 13,104 & 0.52$\pm{0.01}$ &  3.51\\
28496 & 2023 Aug 27 & 129 &  14,210 & 0.58$\pm{0.01}$ & 4.00 \\
28049 & 2023 Oct 20 & 183 & 9,244 & 1.37$\pm{0.02}$ & 9.18 \\
28987 & 2023 Oct 21 & 184 & 8,732 & 1.41$\pm{0.02}$ & 9.42 \\
28988 & 2023 Oct 22 & 185 & 10,768 &1.36$\pm{0.02}$ & 9.11 \\
\enddata
\tablecomments{Exposure length is the  "total on time" minus "dead time". The net (background corrected) count rate is derived from averaging the $\pm{1}$ order event files in the 0.2-2 keV range.}
\end{deluxetable}
\section{Line identifications}
\label{sec:aug_obs} 
\begin{figure}[h!]
 \centering 
 \includegraphics[width=0.98\textwidth]{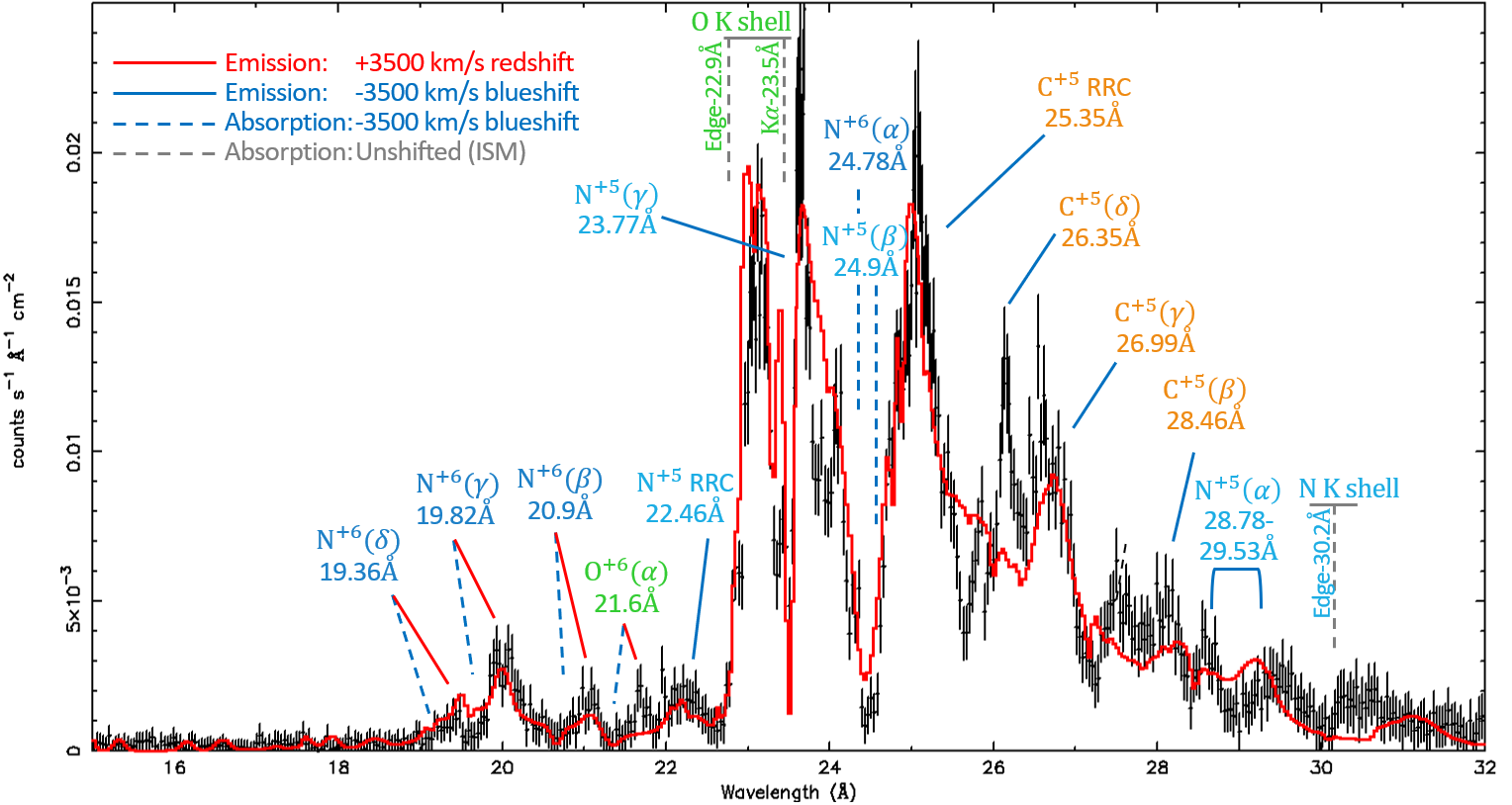}
 
 \caption{Averaged spectrum (two observations) of \sco\ with the \chletg\ 128 days after the optical peak (Obs. ID 28048,28496). The spectrum features an SSS continuum (Sec. \ref{sec:model}), with blueshifted absorption lines of N$^{+5}$, N$^{+6}$, O$^{+6}$, marked by blue dashed lines, and neutral absorption edges of N and O, marked by gray dashed lines. A bright narrow RRC of C$^{+5}$ is observed at 25.35\AA, and the much fainter N$^{+5}$ RRC is at 22.46\AA. The blueshifted CX emission series of C$^{+5}$ and N$^{+5}$, marked by blue solid lines, are observed with intense $\gamma$ lines at 26.99\AA\ and 23.77\AA, respectively. A redshifted CX component is indicated by the N$^{+6}$ series and marked with red solid lines. The total spectral model, marked by the red curve, includes unmarked Ar absorption lines at 24.2\AA\ (Ar$^{+12}$), 26.28\AA, 26.64\AA, 27.44\AA\ (Ar$^{+13}$), and 23.5\AA, 25.9\AA\ (Ar$^{+14}$) that are absent from the atmosphere model and were added as a hot plasma component (Sec.\,\ref{sec:aug_meas}). All quoted wavelengths are in the rest frame.}
 \label{fig:aug_id}
\end{figure}

Fig.\,\ref{fig:aug_id} shows the d128 spectrum (average of two observations) with the ionic identifications. The spectrum shows both absorption and emission features. For nitrogen in absorption, we identify the line series of highly-ionized N$^{+5}$ and N$^{+6}$. In emission, we identify the blueshifted N$^{+5}$ series and RRC, and redshifted series N$^{+6}$, which both feature an enhanced $\gamma$ line ($n=4\rightarrow1$ transition). For oxygen, we identify the O$^{+6}\,\rm{He}\alpha$ absorption and emission lines. For carbon, we identify the C$^{+5}$ Lyman series and the RRC, which hides the much weaker Lyman\,$\alpha$ line of the N$^{+6}$ redshifted series. All RRCs are narrow, indicating a cold temperature of a few eV. Unshifted K-shell absorption edges from neutral N and O, with the latter being especially prominent, are ascribed to the cold circumstellar and ISM. The spectrum shown in Fig.\,\ref{fig:aug_id} is the average of the two observations observed on days 128 and 129, but the fit we show in the next section was done on the separate spectra. 

\begin{figure}[h!]
 \centering 
 \includegraphics[width=0.98\textwidth]{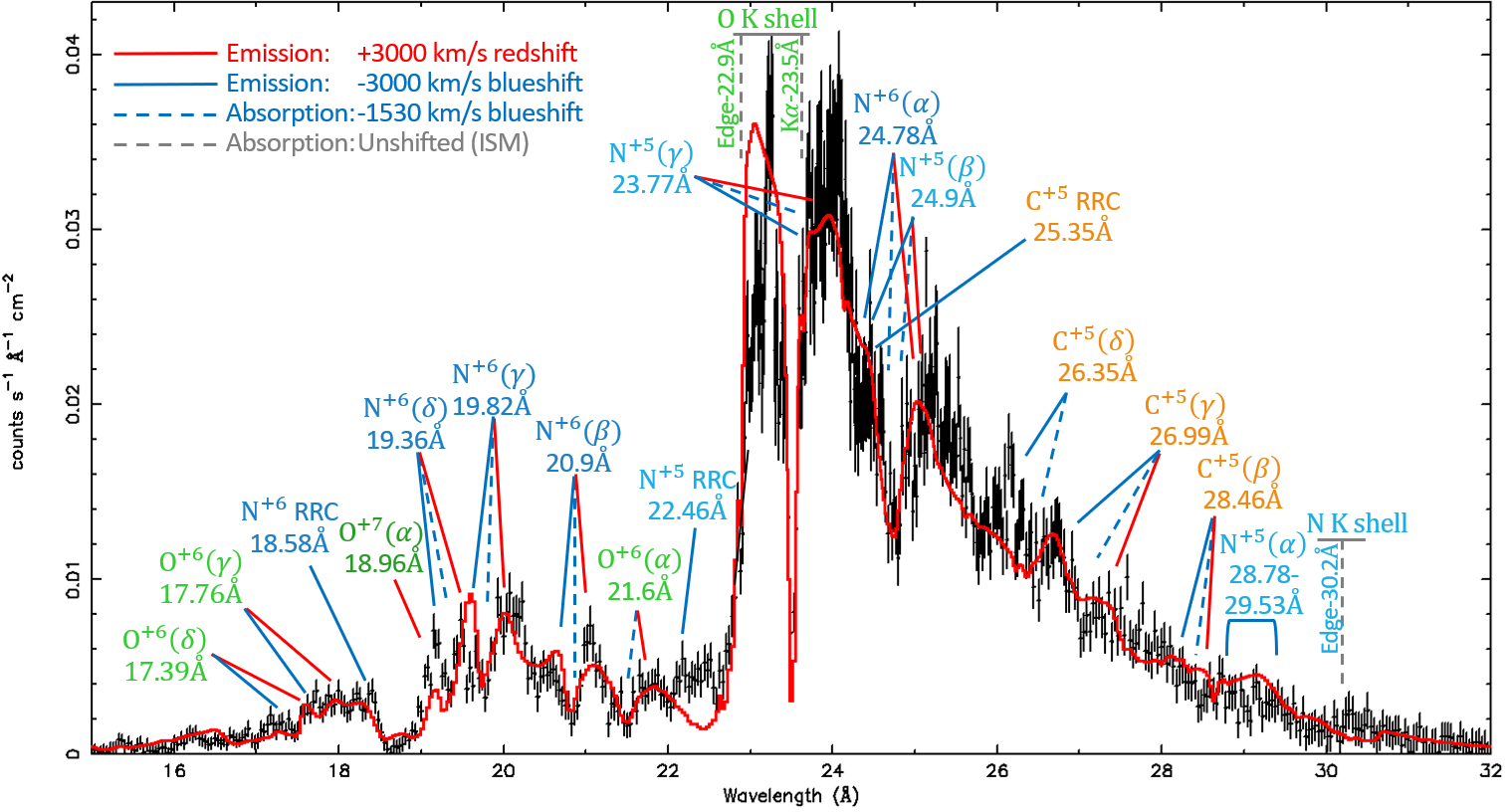}
 
 \caption{Averaged spectrum (three observations) of \sco\ with the \chletg\ 183 days after the optical peak (Obs. ID 228049,28987,28988). The spectrum features an SSS continuum (Sec. \ref{sec:model}), with blueshifted absorption lines of highly ionized C,N,O, marked by blue dashed lines, and neutral absorption edges of N and O, marked by gray dashed lines. The RRCs of C$^{+5}$, N$^{+5}$ and N$^{+6}$ are observed at 25.35\AA, 22.46\AA, and 18.58\AA, respectively. There are two components of CX emission of the same temperature, with an opposite velocity shift of +3000 km/s and -3000 km/s, which are indicated by blue and red solid lines. All quoted wavelengths are in the rest frame.}
 \label{fig:oct_id}
\end{figure}

\label{sec:oct_obs}
 Fig.\,\ref{fig:oct_id} shows the d183 spectrum (average of three observations) that were fitted simultaneously, together with the ionic identifications. The emitted SSS continuum flux is higher in d183, more than double the flux measured in d128, as Table \ref{tab:observations} shows. The atmospheric absorption lines of highly ionized C, N, and O are less blueshifted. We observe an additional N$^{+6}$ RRC, which is somewhat broader than the d128 RRCs, indicating a higher temperature of the cold gas. N$^{+5}$, N$^{+6}$, and C$^{+5}$ emission lines are now visible in both blueshift and redshift. The slower velocity of the atmosphere, and the two emission components, mean that we observe some spectral lines three times. All of the lines are marked in Fig.\,\ref{fig:oct_id} by colored solid (emission) and dashed (absorption) lines. For both figures \ref{fig:aug_id} and \ref{fig:oct_id}, the red curve represents the best fit spectral model,  described in Sec.\,\ref{sec:model} and Table \ref{tb:params}

The accompanying paper (Worley et al.) shows that thermal plasma components in collisional ionization equilibrium, even with line broadening and velocity shifts ('\textit{bvapec}') can not explain the observed emission lines. The reason is the enhanced intensities of high-$n$ emission lines, which are a clear signature of CX emission.
 \subsection{Charge exchange lines} \label{sec:cx_astro}
  Before we proceed, we review
 the properties of CX and how it is applied here. CX emission is a process in which a highly-ionized ion collides with a neutral atom, resulting in transfer of an electron, and thus reducing the ion charge state. The electron is captured into a high-$n$ (principal quantum number) energy level, and the ion eventually decays to the ground state, often producing a high-$n$ emission line.  
The levels populated by CX depend on the collision energy or velocity \citep{Janev1985}. For low velocities, only one or two lines are enhanced. The most populated energy level is given by \citet{Janev1985}:
\begin{equation} \label{eq:cx}
    n=\sqrt{\frac{13.6\,\rm{eV}}{E_{d}}}q \left( 1+\frac{q-1}{\sqrt{2q}} \right)^{-1/2}
\end{equation}
\noindent where $q$ is the ionic charge state, and $E_{d}$ is the ionization potential energy of the donor (cold atom) in eV. 
Spectral models of CX line emission are currently included in astrophysical X-ray spectral packages \citep{Smith2012,Gu2016}. Despite the expectation that CX is ubiquitous for ionized plasma, these models have rarely been tested because the CX line emission is usually too weak and undetected, with a few notable exceptions described below.

CX emission was observed through X-ray emission from the interaction between the ionized solar wind and cold solar system bodies, such as comets \citep{Cravens1997}, and planetary atmospheres (review by \citet{Dennerl2012} and more recently by \citet{Dunn2020}). Both X-ray photometry and spectroscopy were used to study CX in those cases. X-ray images of comets and planets in our solar system show an X-ray foreground which is ascribed to CX emission. Emission lines of C, N, O, and Ne were observed and best fitted to CX emission models \citep{Wolk2009}. 
Outside the solar system, markers of CX are rare and often ambiguous, with only a few cases based on high-resolution spectra. To the best of our knowledge, the only direct evidence for high-n CX emission was recently reported in the grating spectra of nova YZ Ret 2020 \citep{Mitrani2024} and in the microcalorimeter spectrum of supernovae remnant N132D \citep{Gu2025}.

\xrgs\ and \chletg\ spectra of YZ Ret 2020 featured narrow radiative recombination continua (RRCs), doubly-excited satellite lines, and CX emission series, all of which were interpreted as evidence of the interaction between hot ions and cold gas. 
The most conspicuous emission feature in the YZ Ret spectrum is the C$^{+4}$ line series up to $n$=35. This broad population distribution of the high-$n$ atomic levels indicates a high collision velocity or energy \citep{Gu2016}. The XRISM spectrum of N132D features an $n=8,9\rightarrow1$ emission lines of He-like Si. Laboratory colliding beam experiments, indeed produce a high-$n$ population with a broad distribution \citep{Cao2023,Zhu2025}.

Here, we analyze the d128 and d183 LETG spectra of \sco. We specifically model these spectra with a focus on CX signatures. This is also a good opportunity to test and improve the CX spectral models.

\section{Spectral Model}
\label{sec:model}

We construct a composite model, consisting of several spectral components that are available in the XSPEC software package \citep{Arnaud1996}. In this section, we will describe each component, its physical interpretation, and its effect on the observed spectrum. The most prominent feature in the X-ray spectrum is the WD atmosphere. This continuum and its absorption lines are represented here for practical purposes by a hot NLTE stellar atmosphere model \citep{Rauch2003,Rauch2010}, number 003
\footnote{\url{http://astro.uni-tuebingen.de/~rauch/TMAF/flux\_HHeCNONeMgSiS\_gen.html}}
, the two free parameters being temperature and line redshift.  
This static atmospheric model is an approximation for X-ray emitting regions of the outflowing atmosphere of \sco . Another limitation of these models is their assumed abundances. For example, they have no Ar, which we do see in the spectra. Having said that, the models do produce the continuum and the absorption lines. We assume that the temperature uncertainty of the fitted value is half of the step in the model grid.

We fit the neutral absorption edges of O and N with two components of a cold absorber model. The ISM absorption is modeled with the XSPEC model '\textit{tbabs}' and the column density is fixed to the measured value of $N_H=4.4\times10^{21}$\,cm$^{-2}$ \citep{HI4PI2016} and solar elemental abundances. We add a '\textit{tbvarabs}' component which enables fitting the elemental abundances of C, N, O and represents neutral absorption in the circumstellar medium that could be composed of ISM and non-solar nova ejecta. We note that the neutral N K-shell absorption edge in the model is at 30.6\AA\ whereas the correct wavelength is 30.2\AA, and the K$\alpha$ line at 31.6\AA\ is absent. However, for O both are included at their correct wavelengths.

We used XSTAR's \textit{'hotabs'} model \citep{Bautista2001} to model the additional ionized argon absorption lines that are observed but are absent in the atmosphere model. It is used only to indicate these lines in the spectrum, and does not represent an additional hot absorber. 

 We identified narrow RRCs and CX lines, which are the spectral signature of an interaction between shock-heated gas and a cold medium. We model the narrow RRCs with '\textit{vrnei}', which is a recombining non-equilibrium ionization (RNEI) plasma that cooled from temperature $kT_{init}$ to $kT$, with a recombination timescale $\tau=n_e t$ in s\,cm$^{-3}$, where $n_e$ is the electron density and $t$ is time. We suggest that in \sco\ the two temperatures do not represent cooling but the mixing of hot and cold gas. We implement CX by the '\textit{acx}' model \citep{Smith2012}, which produces enhanced intensities of specific high-$n$ lines. All of the emission features are broadened and require further Doppler-broadening ('\textit{gsmooth}'). The full model, described with XSPEC commands is:
 \begin{equation}
    \label{eq:model}
   tbabs(fixed)\times tbvarabs \times [Rauch(003)+gsmooth\times(vrnei+acx(blue)+acx(red))]
\end{equation}
where '\textit{acx(red)}' and '\textit{acx(blue)}' refer to two '\textit{acx}' components with equal and opposite velocity shifts. The best fitted parameters are listed in Table\,\ref{tb:params}. All uncertainties quoted in this paper represent the 90\% statistical confidence. 

\begin{deluxetable}{cccccccc}
\tabletypesize{\footnotesize}
\tablewidth{0pt}
\tablecaption{Best-fit spectral parameters.}
\tablehead{
\colhead{Observation} & \colhead{Component} & \colhead{kT} & \colhead{Velocity} & \colhead{$\sigma_{\nu}$} & \colhead{N$_\mathrm{H}$} & \colhead{N/C} & \colhead{O/C} \\
\colhead{} & \colhead{} & \colhead{[eV]} & \colhead{[km\,s$^{-1}$]} & \colhead{[km\,s$^{-1}$]} & \colhead{[$10^{22}$\,cm$^{-2}$]} & \colhead{[solar abund.]} & \colhead{[solar abund.]}
}
\startdata
\underline{d128$^{(f)}$} & tbabs(fixed) & \nodata & 0 & \nodata & 0.44 & 1 & 1 \\
Obs. ID: & tbvarabs & \nodata & 0 & \nodata & $0.227\pm0.003^{(c2)}$ & $3.5\pm0.5^{(c3)}$ & $2.5\pm0.1^{(c4)}$ \\
28048,28496 & Atmosphere$^{(d)}$ & $64\pm2$ & $-3500\pm150^{(b1)}$ & \nodata & \nodata & \nodata & \nodata \\
{} & vrnei$^{(e)}$ & $86\pm7\rightarrow1.3\pm0.3$ & $-3500\pm150^{(b1)}$ & $^{(c1)}$ & \nodata & $^{(b3)}$ & $^{(b5)}$ \\
{} & CX (blue) & $83\pm3$ & $-3500\pm150^{(b1)}$ & $900\pm100^{(c1)}$ & \nodata & $<2.3^{(b3)}$ & $0.07\pm0.02^{(b5)}$ \\
{} & CX (red) & $172\pm6$ & $+3500\pm150^{(b1)}$ & $^{(c1)}$ & \nodata & $^{(b3)}$ & $^{(b5)}$ \\
\underline{d183$^{(g)}$} & tbabs(fixed) & \nodata & 0 & \nodata & 0.44 & 1 & 1 \\
Obs. ID: & tbvarabs & \nodata & 0 & \nodata & $0.227\pm0.003^{(c2)}$ & $3.5\pm0.5^{(c3)}$ & $2.5\pm0.1^{(c4)}$ \\
28049,28987,28988 & Atmosphere$^{(d)}$ & $66\pm2$ & $-1530\pm30$ & \nodata & \nodata & \nodata & \nodata \\
{} & vrnei$^{(e)}$ & $112\pm2\rightarrow26\pm1$ & $-3000\pm100^{(b2)}$ & $^{(c1)}$ & \nodata & $^{(b4)}$ & $^{(b6)}$ \\
{} & CX (blue) & $128\pm4$ & $-3000\pm100^{(b2)}$ & $900\pm100^{(c1)}$ & \nodata & $2.6\pm0.3^{(b4)}$ & $0.04\pm0.02^{(b6)}$ \\
{} & CX (red) & $128\pm7$ & $+3000\pm100^{(b2)}$ & $^{(c1)}$ & \nodata & $^{(b4)}$ & $^{(b6)}$ \\
\enddata

\tablecomments{\\
($a$) Measured in the range of 0.38–0.83 keV.\\
($b$) Parameters tied together in the fit.\\
($c$) Matching parameters were fixed between observations.\\
($d$) The systematic uncertainties of the atmospheric model are larger than the statistical uncertainties that are reported here.\\
($e$) $\tau=(1.2\pm0.3)\times10^{11}$\,s\,cm$^{-3}$.\\
($f$) C$_\mathrm{stat}$/d.o.f = 4902/2716 = 1.80. Total unabsorbed flux = $(1.8\pm0.1)\times10^{-8}$ [erg\,s$^{-1}$\,cm$^{-2}$].\\
($g$) C$_\mathrm{stat}$/d.o.f = 7351/4795 = 1.53. Total unabsorbed flux = $(2.6\pm0.2)\times10^{-8}$ [erg\,s$^{-1}$\,cm$^{-2}$].\\
}
\end{deluxetable}

\subsection{Day 128 model}
\label{sec:aug_meas}
\begin{figure}[h!]
 \includegraphics[width=0.5\textwidth]{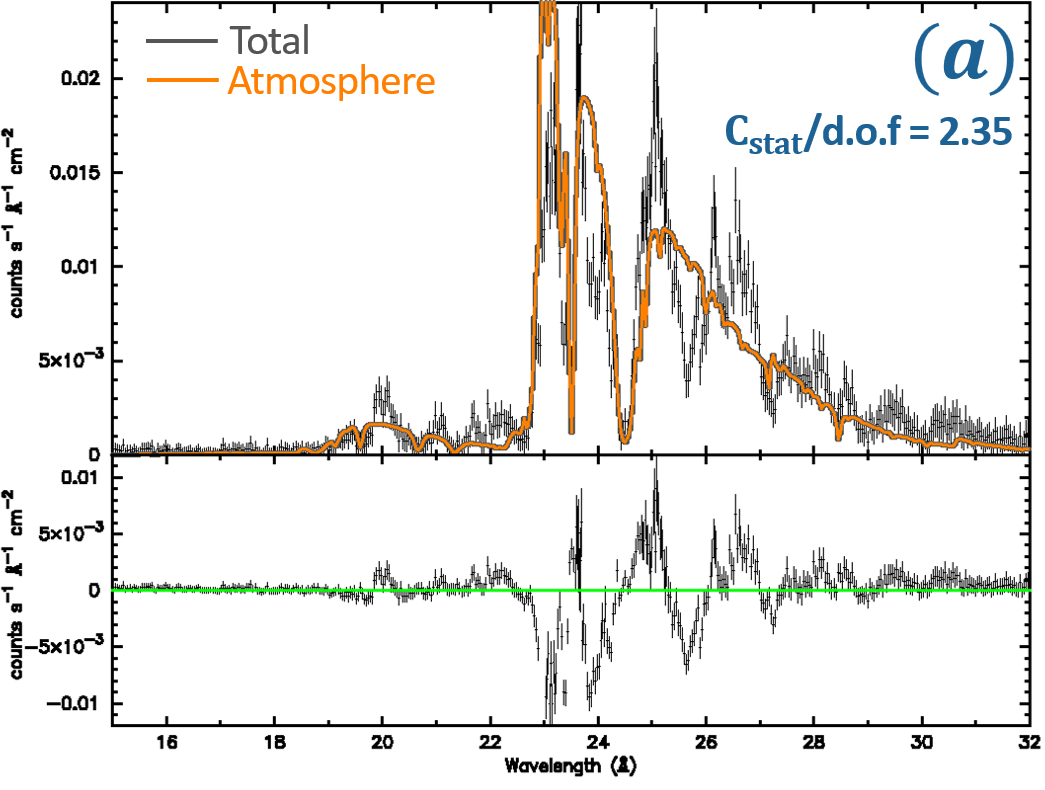}
 \includegraphics[width=0.5\textwidth]{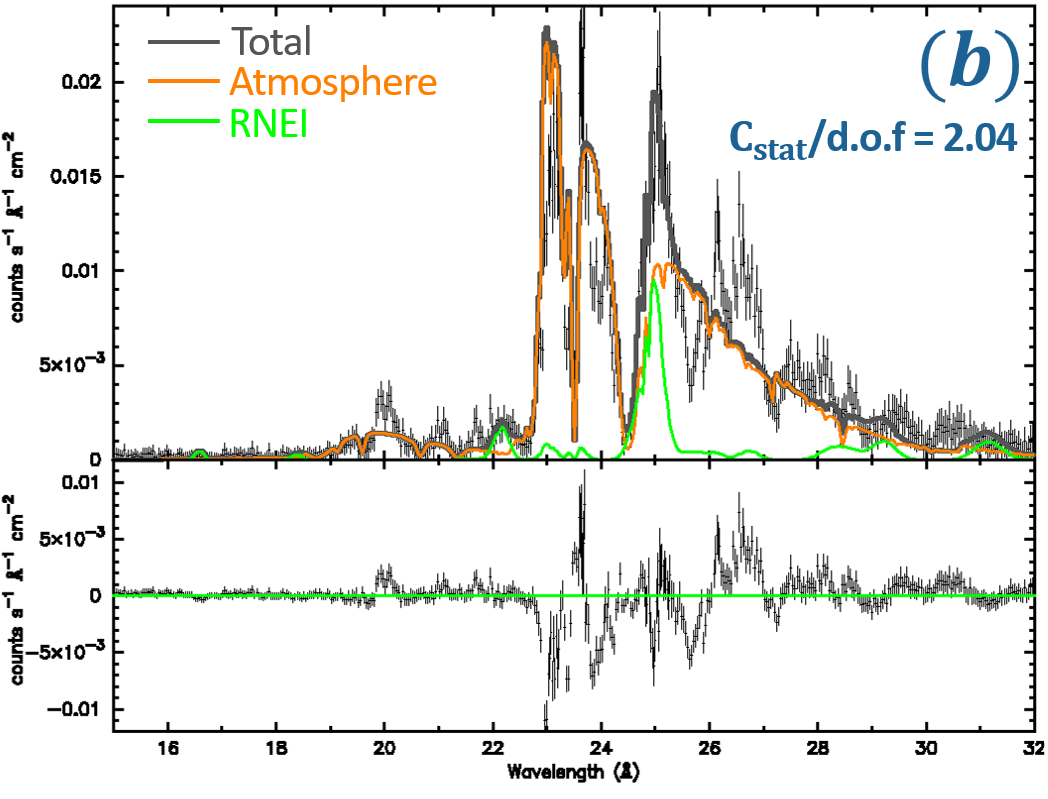}
 \includegraphics[width=0.5\textwidth]{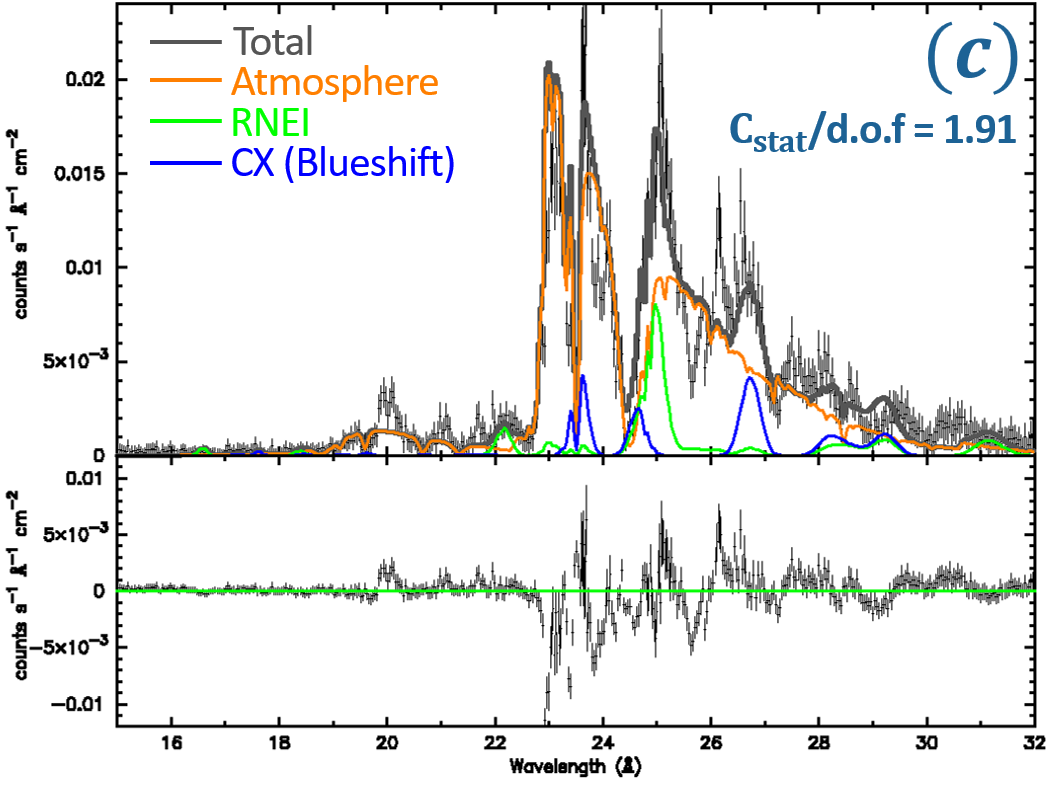}
 \includegraphics[width=0.5\textwidth]{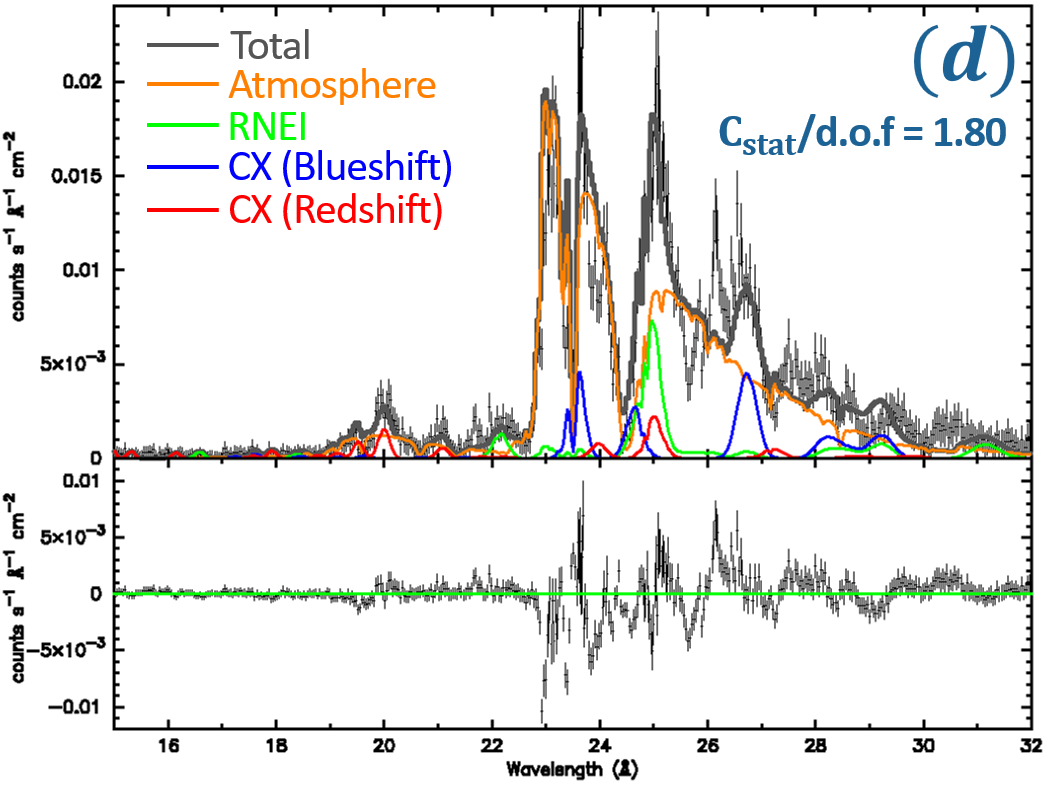}
 \caption{Each panel shows an additional component in the d128 fit, the goodness of fit (C$_{stat}$/d.o.f), and the residuals (lower part of each panel). Panel (a) is the Rauch atmosphere model, with cold absorbers and the approximate Ar lines added with \textit{'hotabs'}. Panel (b) shows the prominent C$^{+5}$ RRC at 25.35\AA, and N$^{+5}$ RRC at 22.46\AA, produced by '\textit{rnei}' plasma model. Panel (c) shows the blueshifted CX emission, mainly indicated by intense $\gamma$ lines of C$^{+5}$ and N$^{+5}$, observed on top of the continuum. Panel (d) shows the redshifted CX emission, highlighted by the highly ionized N$^{+6}$ series.} 
 \label{fig:aug_fit}
\end{figure}

Fig.\,\ref{fig:aug_fit} shows each component in the full model of d128. Panel (a) shows the WD atmosphere model, with a temperature of $kT=64\pm2$\,eV. Since there is no Ar in this model, we added a \textit{'hotabs'} component of only Ar to approximately match the observed Ar absorption lines at 24.2\AA\ (Ar$^{+12}$), 26.28\AA, 26.64\AA, 27.44\AA\ (Ar$^{+13}$), and 23.5\AA, 25.9\AA\ (Ar$^{+14}$). Because the 'hotabs' assumes a low density plasma, the best-fitted temperature is higher than that of the atmosphere, to which we ascribe no physical meaning.
In panel (b), we added the recombining plasma model that produces the two RRCs of C$^{+5}$ and N$^{+5}$. It corresponds to a mixing of $kT=86\pm7$\,eV hot plasma and $kT=1.3\pm0.3$\,eV cold gas, with $\tau=(1.2\pm0.3)\times10^{11}$\,s\,cm$^{-3}$. In panel (c), we added the blueshifted CX emission model with $kT=83\pm3$\,eV. It produces the series of C$^{+5}$ and N$^{+5}$, with an intense $\gamma$ line at 26.99\AA, and 23.77\AA, respectively. A bright C$^{+5}$ Lyman\,$\delta$ is prominent in the data but not in the model. Finally, in panel (d), we added the redshifted CX emission model with $kT=172\pm6$\,eV, which produces the N$^{+6}$ Lyman series. The different temperatures between the CX components could be due to a disparity in density that governs the heating and cooling timescales.

All emission components are broadened by $\sigma_{\nu}=900$\,km/s, and blueshifted by $v=-3500\pm150$\,km/s, except '\textit{acx(red)}' which has an opposite velocity shift of $v=+3500\pm150$\,km/s. We note that fitting the circumstellar cold absorber requires overabundance of N and O. The emission components require also overabundance of N, but underabundance of O, indicated by its weak emission lines. A puzzling line is observed at 21.7\AA, which may be a redshifted O$^{+6}$ He$\alpha$ triplet (21.60-22.10\AA), but this would be inconsistent with the overall lack of O in the spectrum. The fit was implemented with Cash statistics \citep[Cstat,][]{Cash79}, and we obtained a goodness of fit of Cstat/d.o.f = 4902/2716 = 1.80. 

\subsection{Day 183 model}
\label{sec:oct_meas}

\begin{figure}[h!]
 
 \includegraphics[width=0.505\textwidth]{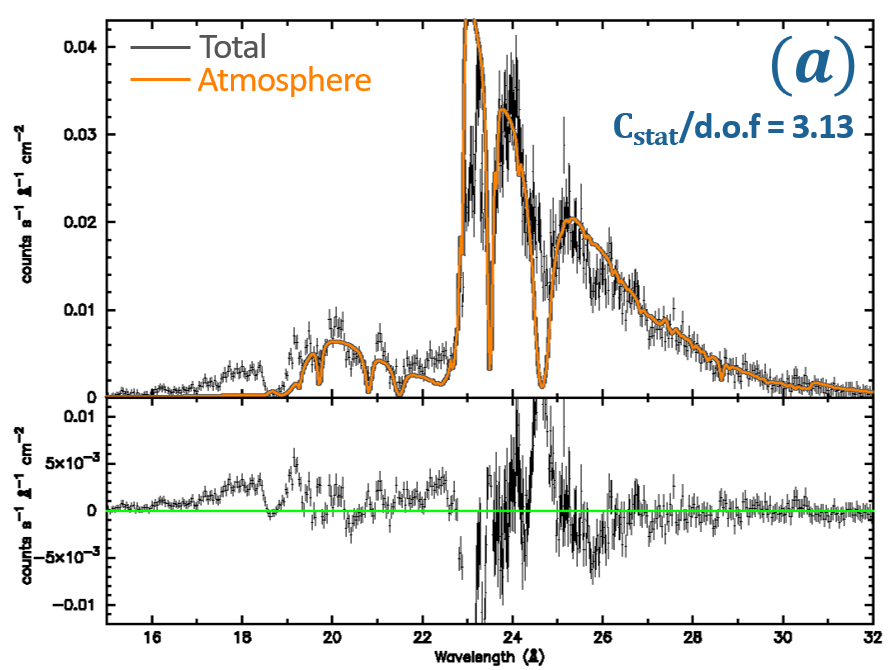}
 \includegraphics[width=0.495\textwidth]{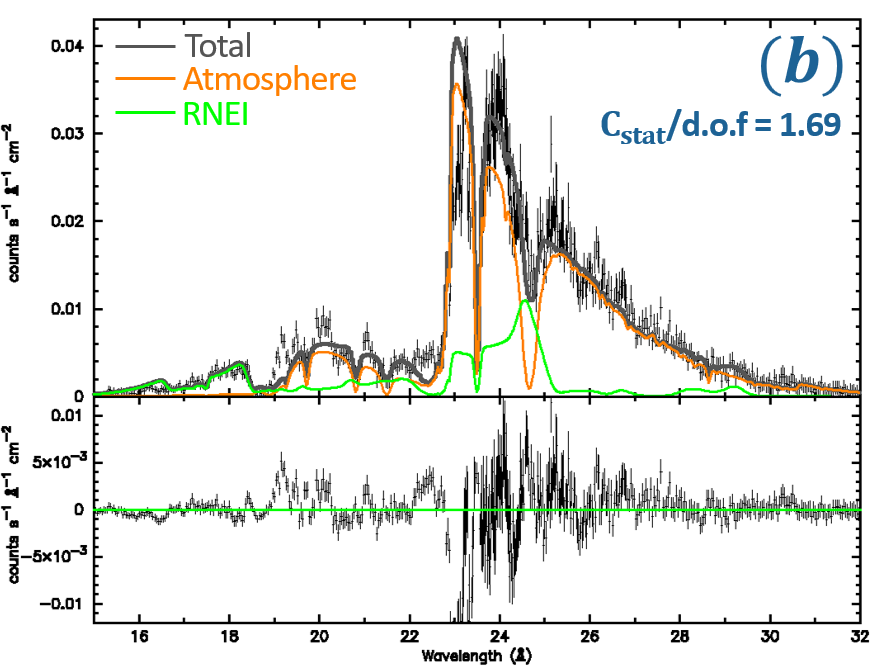}
 \includegraphics[width=0.5\textwidth]{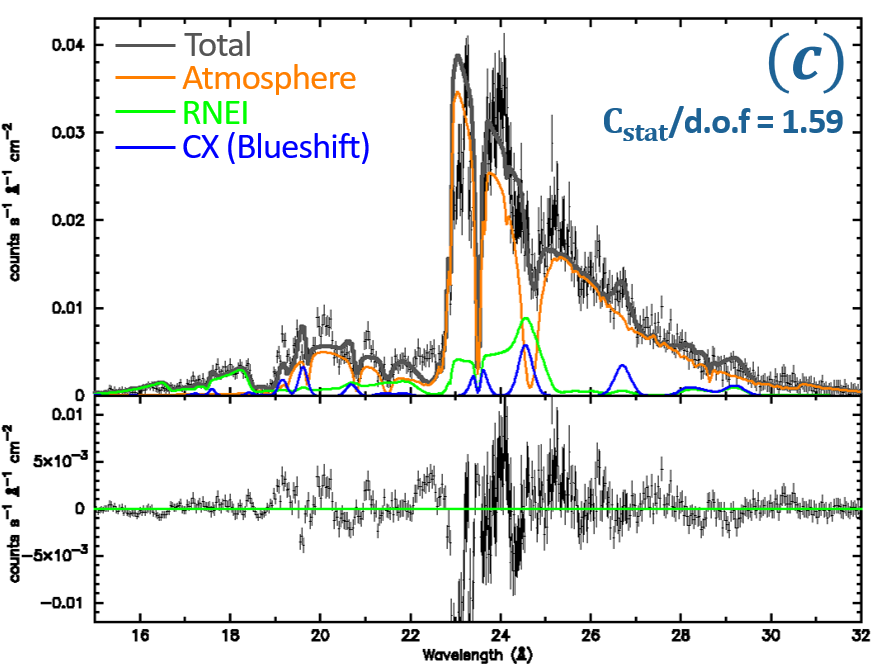}
 \includegraphics[width=0.5\textwidth]{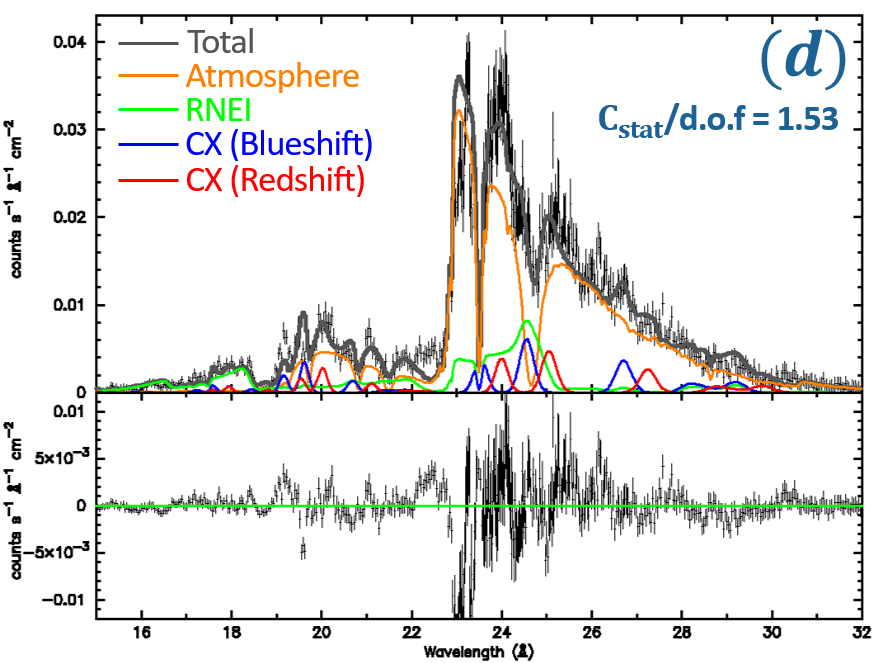}
 \caption{Each panel shows an additional component in the d183 fit, the goodness of fit (C$_{stat}$/d.o.f), and the residuals (lower part of each panel). Panel (a) is the Rauch atmosphere model, with cold absorbers. Panel (b) shows the C$^{+5}$, N$^{+5}$, and N$^{+6}$ produced by RNEI plasma model. Panel (c) and (d) show the blue- and red-shifted CX emission, observed on top of the continuum.} 
 \label{fig:oct_fit}
\end{figure}

Fig.\,\ref{fig:oct_fit} shows each component in the full model of d183. Panel (a) shows the Rauch model of the WD atmosphere, with a temperature of  $kT=66\pm2$\,eV, matching to the d128 value. Unlike d128, no Ar lines are observed in the spectrum. In panel (b), we added the recombining plasma model that produces the RRCs of C$^{+5}$, N$^{+5}$, and N$^{+6}$. It corresponds to a mixing of $kT=112\pm2$\,eV hot plasma and $kT=26\pm1$\,eV cold gas, indicated by broader RRCs than d128. In panel (c), we added the blueshifted CX emission model which shows emission lines of the ions at a higher ionization degree, produced by a temperature $kT=128\pm4$\,eV. In panel (d), we included the redshifted CX model with $kT=128\pm7$\,eV, matching with the blueshifted component temperature. Both CX components produce the C$^{+5}$, N$^{+5}$ and N$^{+6}$ emission series with a prominent enhanced $\gamma$ emission lines, but underestimate the intensity of the $\delta$ lines. O$^{+6}$ He$\gamma$ and He$\delta$ are weak owing to the low O abundance, but can be observed at $\sim18$\AA\ on top of the N$^{+6}$ RRC.

The atmospheric lines are now blueshifted by $v=-1530\pm30$\,km/s, and the recombining plasma and two CX components have an opposite velocity shifts of $\pm3000(\pm100)$\,km/s, which is close to the d128 value. The emission line broadening, and the cold absorber abundances were set to the d128 values. The abundances of the emission components were fitted separately and resulted in similar values to d128. The goodness of fit is Cstat/d.o.f = 7351/4795 = 1.53.

\subsection{CX model comparison with data} 
The strongest indication for CX emission are the enhanced intensities of the C$^{+5}$ Lyman\,$\gamma$ (26.99\AA), N$^{+5}\,\rm{He}\gamma$ (23.77\AA), and N$^{+6}$ Lyman\,$\gamma$ (19.82\AA) lines in both d128 and d183, which correspond to $n=4\rightarrow1$ energy level transition. 
These lines are adequately fitted by the \textit{'acx'} model, and match the prediction of the $n=4$ level being the most populated energy level according to Eq.\,\ref{eq:cx}.
However, \sco\ also shows strong Lyman\,$\delta$ lines, most conspicuously for C$^{+5}$ at 26.35\AA, which the model does not reproduce. A weak Lyman\,$\varepsilon$ of C$^{+5}$ may be present at 26.02\AA, but given the limitations of the atmospheric model, it can not be detected with any certainty.

The collision velocity affects the CX series intensities. The \textit{'acx'} model assumes the low velocity limit. Higher collision velocity will create a broader and more uniform population distribution of the energy levels  \citep{Gu2016,Cao2023}, as observed in our spectra. A CX model that implements this velocity dependence might be able to better match the data. Other effects that can change the population distribution of the ions are CX with excited H or with He. The '\textit{acx}' model does have a He abundance parameter that affects the line intensities. 
The best-fit He abundances are upper limits of $<11\%$ in d128, consistent with the solar abundance (9\%), and $<2\%$ in d183. 
The low He abundance estimated for d183 changed the fit very marginally, therefore in both fits we leave the He abundance to be solar.

\section{Discussion}
\label{sec:discuss}
Based on the premise that the nova X-ray spectrum is composed of two components, an atmospheric continuum with absorption features and emission lines from ionized plasma due to shocked gas, we successfully composed a model that well fits the spectra. During both epochs, the atmosphere temperature is the same at $T\sim750,000$\,K ($kT\sim65$\,eV).  
An  effective temperature of 750,000\,K would be attributed in the theoretical models to a WD of about 1.1 M$_\odot$ \citep{Starrfield2012, Wolf2013}, which is consistent also with the WD mass-dependent duration of the SSS phase, estimated in the accompanying paper (Worley et al.). 
The variability reported in the Swift and NICER monitoring observations in the accompanying paper (Worley et al.) could be explained by varying absorption in the clumpy circumstellar ejecta or rotation of X-ray emitting regions in and out of the line of sight.

On d128, the atmosphere expanded at a velocity of $v=-3500$\,km\,s$^{-1}$. An X-ray bipolar outflow indicated by red- and blue-shifted emission components also at $v=\pm3500$\,km\,s$^{-1}$ suggest we are observing shocked ejecta originating from the atmosphere. Evidence for a fast outflow was reported in April 2023 by the optical observations of \citet{Shore2023a,Izzo2023,Shore2023b}, which show P-Cygni lines with velocities of up to $-3200$\,km/s. Interestingly, \citet{Woodward2024} reported double-peaked infrared lines in an observation 132 days after the optical peak (in August) of \sco, with slower velocity shifts of $\pm1000$\,km\,s$^{-1}$, corroborating the bipolar outflow scenario. 
They discuss how bipolar outflows are common to novae. In the last 20 years high spatial resolution images including with interferometric methods resolve these outflows. \citet{Chomiuk2021} discuss similar evidence from radio observations. Old optical remnants of novae show bipolar structure a long time after the outburst \citep{Duerbeck1992, Williams1979}. Thus, spectroscopic signatures of a bipolar outflow several months after the onset of the outburst should not come as a total surprise. On the other hand, it is rare to simultaneously observe X-ray blue- and red- shifted lines, which are a unique sign of fast bipolar shocks, which are more extended than the opaque central source. 

The atmospheric absorption features are less blueshifted on d183, with a velocity of only $v=-1530$\,km\,s$^{-1}$ compared to $v=-3500$\,km\,s$^{-1}$ on d128. The emission line components slightly slowed down from $v=\pm3500$\,km\,s$^{-1}$ on d128, to $v=\pm3000$\,km\,s$^{-1}$ on d183. This indicates a physical separation between the WD atmosphere and shocked gas. The two bipolar components are now at the same temperature of $kT=128$\,eV as opposed to d128, indicating shock heating and cooling is ongoing on months timescales.


\section{Conclusions} \label{sec:conclu}

\begin{itemize}
    \item We successfully implemented a composite spectral model of an absorbed continuum, recombining plasma emission, and CX emission to fit all spectra of \sco.
    \item The continuum emission is ascribed to the WD atmosphere at a temperature of $T \simeq750,000$\,K ($kT \simeq65$\,eV), outflowing at $v\simeq-3500$\,km\,s$^{-1}$ on d128, and slowing down to $v\simeq-1500$\,km\,s$^{-1}$ on d183.
    \item The spectral line emission is ascribed to plasma heated by a shock in the nova ejecta. The narrow RRCs and CX emission are evidence of mixing between the shock-heated plasma and cold gas in the surrounding area. The RRCs are produced by hot ions recombining with cold electrons: $kT_e =1.3$\,eV on d128, and $kT_e =26$\,eV on d183. 
    The CX high-$n$ emission lines are uniquely due to highly ionized ions colliding with neutral atoms. The hot plasma temperature is measured to be between $kT=83-172$\,eV. 
    \item The preferred interpretation of the two CX kinematic components with opposite velocities of $v=\pm3000$\,km\,s$^{-1}$ is a bipolar outflow.
    \item Given the evidence for CX in YZ Ret and \sco, other grating spectra of novae in the Chandra and XMM-Newton archives should be analyzed, seeking these subtle signatures of CX.
\end{itemize}

\bigskip
We thank the referee for a meticulous report that considerably improved the presentation of our results. S.M. and E.B. were partly supported by a grant from The Israeli Institute for Fusion Research (IIFR). As principal investigator of the {\sl Chandra} project, M.O.was funded by an award of
 the Smithsonian Observatory to the University of Wisconsin. This paper employs a list of Chandra datasets, obtained by the Chandra X-ray Observatory (P.I. Orio), contained in the Chandra Data Collection ~\dataset[DOI: 10.25574/cdc.413]{https://doi.org/10.25574/cdc.413}.

\bibliography{main}{}

\begin{thebibliography}{}
\expandafter\ifx\csname natexlab\endcsname\relax\def\natexlab#1{#1}\fi
\providecommand{\url}[1]{\href{#1}{#1}}
\providecommand{\dodoi}[1]{doi:~\href{http://doi.org/#1}{\nolinkurl{#1}}}
\providecommand{\doeprint}[1]{\href{http://ascl.net/#1}{\nolinkurl{http://ascl.net/#1}}}
\providecommand{\doarXiv}[1]{\href{https://arxiv.org/abs/#1}{\nolinkurl{https://arxiv.org/abs/#1}}}

\bibitem[{{Arnaud}(1996)}]{Arnaud1996}
{Arnaud}, K.~A. 1996, in Astronomical Society of the Pacific Conference Series, Vol. 101, Astronomical Data Analysis Software and Systems V, ed. G.~H. {Jacoby} \& J.~{Barnes}, 17

\bibitem[{{Bautista} \& {Kallman}(2001)}]{Bautista2001}
{Bautista}, M.~A., \& {Kallman}, T.~R. 2001, \apjs, 134, 139, \dodoi{10.1086/320363}

\bibitem[{{Cao} {et~al.}(2023){Cao}, {Meng}, {Gao}, {Zhang}, {Zhang}, {Yan}, {Zhu}, {Wang}, {Ma}, {Ren}, {Xia}, {Guo}, {Zhang}, {Lin}, {Xu}, {Wei}, \& {Ma}}]{Cao2023}
{Cao}, T., {Meng}, T., {Gao}, Y., {et~al.} 2023, \apjs, 266, 20, \dodoi{10.3847/1538-4365/accba2}

\bibitem[{{Cash}(1979)}]{Cash79}
{Cash}, W. 1979, \apj, 228, 939, \dodoi{10.1086/156922}

\bibitem[{{Cheung}(2023)}]{Cheung2023}
{Cheung}, C.~C. 2023, The Astronomer's Telegram, 16002, 1

\bibitem[{{Chomiuk} {et~al.}(2021){Chomiuk}, {Linford}, {Aydi}, {Bannister}, {Krauss}, {Mioduszewski}, {Mukai}, {Nelson}, {Rupen}, {Ryder}, {Sokoloski}, {Sokolovsky}, {Strader}, {Filipovi{\'c}}, {Finzell}, {Kawash}, {Kool}, {Metzger}, {Nyamai}, {Ribeiro}, {Roy}, {Urquhart}, \& {Weston}}]{Chomiuk2021}
{Chomiuk}, L., {Linford}, J.~D., {Aydi}, E., {et~al.} 2021, \apjs, 257, 49, \dodoi{10.3847/1538-4365/ac24ab}

\bibitem[{{Cravens}(1997)}]{Cravens1997}
{Cravens}, T.~E. 1997, \grl, 24, 105, \dodoi{10.1029/96GL03780}

\bibitem[{{Dennerl} {et~al.}(2012){Dennerl}, {Lisse}, {Bhardwaj}, {Christian}, {Wolk}, {Bodewits}, {Zurbuchen}, {Combi}, \& {Lepri}}]{Dennerl2012}
{Dennerl}, K., {Lisse}, C.~M., {Bhardwaj}, A., {et~al.} 2012, Astronomische Nachrichten, 333, 324, \dodoi{10.1002/asna.201211663}

\bibitem[{{Drake} {et~al.}(2016){Drake}, {Delgado}, {Laming}, {Starrfield}, {Kashyap}, {Orlando}, {Page}, {Hernanz}, {Ness}, {Gehrz}, {van Rossum}, \& {Woodward}}]{Drake2016}
{Drake}, J.~J., {Delgado}, L., {Laming}, J.~M., {et~al.} 2016, \apj, 825, 95, \dodoi{10.3847/0004-637X/825/2/95}

\bibitem[{{Dunn} {et~al.}(2020){Dunn}, {Branduardi-Raymont}, {Carter-Cortez}, {Campbell}, {Elsner}, {Ness}, {Gladstone}, {Ford}, {Yao}, {Rodriguez}, {Clark}, {Paranicas}, {Foster}, {Baker}, {Gray}, {Badman}, {Ray}, {Bunce}, {Snios}, {Jackman}, {Rae}, {Kraft}, {Rymer}, {Lathia}, \& {Achilleos}}]{Dunn2020}
{Dunn}, W.~R., {Branduardi-Raymont}, G., {Carter-Cortez}, V., {et~al.} 2020, Journal of Geophysical Research (Space Physics), 125, e27219, \dodoi{10.1029/2019JA027219}

\bibitem[{{Evans} {et~al.}(1992){Evans}, {Bode}, {Duerbeck}, \& {Seitter}}]{Duerbeck1992}
{Evans}, A., {Bode}, M.~F., {Duerbeck}, H.~W., \& {Seitter}, W.~C. 1992, \mnras, 258, 7P, \dodoi{10.1093/mnras/258.1.7P}

\bibitem[{{Fruscione} {et~al.}(2006){Fruscione}, {McDowell}, {Allen}, {Brickhouse}, {Burke}, {Davis}, {Durham}, {Elvis}, {Galle}, {Harris}, {Huenemoerder}, {Houck}, {Ishibashi}, {Karovska}, {Nicastro}, {Noble}, {Nowak}, {Primini}, {Siemiginowska}, {Smith}, \& {Wise}}]{Fruscione2006}
{Fruscione}, A., {McDowell}, J.~C., {Allen}, G.~E., {et~al.} 2006, in Society of Photo-Optical Instrumentation Engineers (SPIE) Conference Series, Vol. 6270, Observatory Operations: Strategies, Processes, and Systems, ed. D.~R. {Silva} \& R.~E. {Doxsey}, 62701V, \dodoi{10.1117/12.671760}

\bibitem[{{Gu} {et~al.}(2016){Gu}, {Kaastra}, \& {Raassen}}]{Gu2016}
{Gu}, L., {Kaastra}, J., \& {Raassen}, A.~J.~J. 2016, \aap, 588, A52, \dodoi{10.1051/0004-6361/201527615}

\bibitem[{{Gu} {et~al.}(2025){Gu}, {Yamaguchi}, {Foster}, {Katsuda}, {Uchida}, {Sawada}, {Porter}, {Williams}, {Petre}, {Bamba}, {Terada}, {Agarwal}, {Decourchelle}, {Guainazzi}, {Kelley}, {Kilbourne}, {Loewenstein}, {Matsumoto}, {Miller}, {Ohshiro}, {Plucinsky}, {Suzuki}, {Tashiro}, {Vink}, {Ezoe}, {Behar}, \& {Smith}}]{Gu2025}
{Gu}, L., {Yamaguchi}, H., {Foster}, A., {et~al.} 2025, arXiv e-prints, arXiv:2504.03223, \dodoi{10.48550/arXiv.2504.03223}

\bibitem[{{HI4PI Collaboration} {et~al.}(2016){HI4PI Collaboration}, {Ben Bekhti}, {Fl{\"o}er}, {Keller}, {Kerp}, {Lenz}, {Winkel}, {Bailin}, {Calabretta}, {Dedes}, {Ford}, {Gibson}, {Haud}, {Janowiecki}, {Kalberla}, {Lockman}, {McClure-Griffiths}, {Murphy}, {Nakanishi}, {Pisano}, \& {Staveley-Smith}}]{HI4PI2016}
{HI4PI Collaboration}, {Ben Bekhti}, N., {Fl{\"o}er}, L., {et~al.} 2016, \aap, 594, A116, \dodoi{10.1051/0004-6361/201629178}

\bibitem[{{Izzo} {et~al.}(2023){Izzo}, {Della Valle}, {Woodward}, {Molaro}, {Starrfield}, {Wagner}, {Guido}, {Cecconi}, \& {Padilla-Torres}}]{Izzo2023}
{Izzo}, L., {Della Valle}, M., {Woodward}, C.~E., {et~al.} 2023, The Astronomer's Telegram, 16007, 1

\bibitem[{Janev \& Winter(1985)}]{Janev1985}
Janev, R., \& Winter, H. 1985, Physics Reports, 117, 265, \dodoi{https://doi.org/10.1016/0370-1573(85)90118-8}

\bibitem[{{Mitrani} {et~al.}(2024){Mitrani}, {Behar}, {Drake}, {Orio}, {Page}, {Canton}, {Ness}, \& {Sokolovsky}}]{Mitrani2024}
{Mitrani}, S., {Behar}, E., {Drake}, J.~J., {et~al.} 2024, \apj, 970, 54, \dodoi{10.3847/1538-4357/ad4a64}

\bibitem[{{Nelson} {et~al.}(2008){Nelson}, {Orio}, {Cassinelli}, {Still}, {Leibowitz}, \& {Mucciarelli}}]{Nelson2008}
{Nelson}, T., {Orio}, M., {Cassinelli}, J.~P., {et~al.} 2008, \apj, 673, 1067, \dodoi{10.1086/524054}

\bibitem[{{Ness} {et~al.}(2013){Ness}, {Osborne}, {Henze}, {Dobrotka}, {Drake}, {Ribeiro}, {Starrfield}, {Kuulkers}, {Behar}, {Hernanz}, {Schwarz}, {Page}, {Beardmore}, \& {Bode}}]{Ness2013}
{Ness}, J.~U., {Osborne}, J.~P., {Henze}, M., {et~al.} 2013, \aap, 559, A50, \dodoi{10.1051/0004-6361/201322415}

\bibitem[{{Ness} {et~al.}(2023){Ness}, {Beardmore}, {Bode}, {Darnley}, {Dobrotka}, {Drake}, {Magdolen}, {Munari}, {Osborne}, {Orio}, {Page}, \& {Starrfield}}]{Ness2023}
{Ness}, J.~U., {Beardmore}, A.~P., {Bode}, M.~F., {et~al.} 2023, \aap, 670, A131, \dodoi{10.1051/0004-6361/202245269}

\bibitem[{{Orio}(2012)}]{Orio2012}
{Orio}, M. 2012, Bulletin of the Astronomical Society of India, 40, 333, \dodoi{10.48550/arXiv.1210.4331}

\bibitem[{{Orio} {et~al.}(2020){Orio}, {Drake}, {Ness}, {Behar}, {Luna}, {Darnley}, {Gallagher}, {Gehrz}, {Kuin}, {Mikolajewska}, {Ospina}, {Page}, {Poggiani}, {Starrfield}, {Williams}, \& {Woodward}}]{Orio2020}
{Orio}, M., {Drake}, J.~J., {Ness}, J.~U., {et~al.} 2020, \apj, 895, 80, \dodoi{10.3847/1538-4357/ab8c4d}

\bibitem[{{Orio} {et~al.}(2022){Orio}, {Gendreau}, {Giese}, {Luna}, {Magdolen}, {Pei}, {Sun}, {Behar}, {Dobrotka}, {Mikolajewska}, {Pasham}, \& {Strohmayer}}]{Orio2022}
{Orio}, M., {Gendreau}, K., {Giese}, M., {et~al.} 2022, \apj, 932, 45, \dodoi{10.3847/1538-4357/ac63be}

\bibitem[{{Peretz} {et~al.}(2016){Peretz}, {Orio}, {Behar}, {Bianchini}, {Gallagher}, {Rauch}, {Tofflemire}, \& {Zemko}}]{Peretz2016}
{Peretz}, U., {Orio}, M., {Behar}, E., {et~al.} 2016, \apj, 829, 2, \dodoi{10.3847/0004-637X/829/1/2}

\bibitem[{{Rauch} {et~al.}(2010){Rauch}, {Orio}, {Gonzales-Riestra}, {Nelson}, {Still}, {Werner}, \& {Wilms}}]{Rauch2010}
{Rauch}, T., {Orio}, M., {Gonzales-Riestra}, R., {et~al.} 2010, \apj, 717, 363, \dodoi{10.1088/0004-637X/717/1/363}

\bibitem[{{Rauch, T.}(2003)}]{Rauch2003}
{Rauch, T.} 2003, A\&A, 403, 709, \dodoi{10.1051/0004-6361:20030412}

\bibitem[{{Shore} {et~al.}(2023{\natexlab{a}}){Shore}, {Charbonnel}, {Le Du}, {Garde}, {Mulato}, \& {Petit}}]{Shore2023a}
{Shore}, S., {Charbonnel}, S., {Le Du}, P., {et~al.} 2023{\natexlab{a}}, The Astronomer's Telegram, 16004, 1

\bibitem[{{Shore} {et~al.}(2023{\natexlab{b}}){Shore}, {Charbonnel}, {Le Du}, {Garde}, {Mulato}, {Petit}, \& {Curry}}]{Shore2023b}
---. 2023{\natexlab{b}}, The Astronomer's Telegram, 16036, 1

\bibitem[{{Smith} {et~al.}(2001){Smith}, {Brickhouse}, {Liedahl}, \& {Raymond}}]{Smith2001}
{Smith}, R.~K., {Brickhouse}, N.~S., {Liedahl}, D.~A., \& {Raymond}, J.~C. 2001, \apjl, 556, L91, \dodoi{10.1086/322992}

\bibitem[{{Smith} {et~al.}(2012){Smith}, {Foster}, \& {Brickhouse}}]{Smith2012}
{Smith}, R.~K., {Foster}, A.~R., \& {Brickhouse}, N.~S. 2012, Astronomische Nachrichten, 333, 301, \dodoi{10.1002/asna.201211673}

\bibitem[{{Sokolovsky} {et~al.}(2023){Sokolovsky}, {Aydi}, {Chomiuk}, {Strader}, {Sokoloski}, {Linford}, {Mukai}, {Buckley}, {Mikolajewska}, {Orio}, {Stanek}, {Kochanek}, {Hambsch}, {Odeh}, {Pearce}, {Romanov}, {Sharp}, {Verveer}, \& {Young}}]{Sokolovsky2023}
{Sokolovsky}, K., {Aydi}, E., {Chomiuk}, L., {et~al.} 2023, The Astronomer's Telegram, 16018, 1

\bibitem[{{Starrfield} {et~al.}(2012){Starrfield}, {Iliadis}, {Timmes}, {Hix}, {Arnett}, {Meakin}, \& {Sparks}}]{Starrfield2012}
{Starrfield}, S., {Iliadis}, C., {Timmes}, F.~X., {et~al.} 2012, Bulletin of the Astronomical Society of India, 40, 419, \dodoi{10.48550/arXiv.1210.6086}

\bibitem[{{Wang} {et~al.}(2024){Wang}, {Yan}, {Takata}, \& {Lin}}]{Wang2024}
{Wang}, H.~H., {Yan}, H.~D., {Takata}, J., \& {Lin}, L.~C.~C. 2024, arXiv e-prints, arXiv:2406.19233, \dodoi{10.48550/arXiv.2406.19233}

\bibitem[{{Williams} \& {Gallagher}(1979)}]{Williams1979}
{Williams}, R.~E., \& {Gallagher}, J.~S. 1979, \apj, 228, 482, \dodoi{10.1086/156869}

\bibitem[{{Wolf} {et~al.}(2013){Wolf}, {Bildsten}, {Brooks}, \& {Paxton}}]{Wolf2013}
{Wolf}, W.~M., {Bildsten}, L., {Brooks}, J., \& {Paxton}, B. 2013, \apj, 777, 136, \dodoi{10.1088/0004-637X/777/2/136}

\bibitem[{{Wolk} {et~al.}(2009){Wolk}, {Lisse}, {Bodewits}, {Christian}, \& {Dennerl}}]{Wolk2009}
{Wolk}, S.~J., {Lisse}, C.~M., {Bodewits}, D., {Christian}, D.~J., \& {Dennerl}, K. 2009, \apj, 694, 1293, \dodoi{10.1088/0004-637X/694/2/1293}

\bibitem[{{Woodward} {et~al.}(2024){Woodward}, {Shaw}, {Starrfield}, {Evans}, \& {Page}}]{Woodward2024}
{Woodward}, C.~E., {Shaw}, G., {Starrfield}, S., {Evans}, A., \& {Page}, K.~L. 2024, \apj, 968, 31, \dodoi{10.3847/1538-4357/ad4097}

\bibitem[{{Zhu} {et~al.}(2025){Zhu}, {Zhang}, {Zhang}, {Mitrani}, {Gu}, {Gao}, {Zhang}, \& {Ma}}]{Zhu2025}
{Zhu}, X.~B., {Zhang}, R.~T., {Zhang}, C.~J., {et~al.} 2025, \apjs, 277, 35, \dodoi{10.3847/1538-4365/adb72a}

\end{thebibliography}
\bibliographystyle{aasjournal}



\end{document}